\documentclass[showpacs,english,twocolumn,pre,superscriptaddress]{revtex4-1}
\usepackage[T1]{fontenc}
\usepackage[latin9]{inputenc}
\usepackage{color}
\usepackage{babel}
\usepackage{units}
\usepackage{textcomp}
\usepackage{amsmath}
\usepackage{amssymb}
\usepackage{graphicx}
\usepackage{setspace}
\usepackage{esint}
\usepackage{ulem}
\usepackage{siunitx}

\usepackage[unicode=true,pdfusetitle,
 bookmarks=true,bookmarksnumbered=false,bookmarksopen=false,
 breaklinks=false,pdfborder={0 0 1},backref=false,colorlinks=true]
 {hyperref}
\hypersetup{linkcolor=blue,urlcolor=blue,citecolor=blue}
        
\definecolor{light-gray}{gray}{0.80}
\begin{document}
\title{Time-delayed nonlocal response induces traveling localized structures}
        
\author{M.G. Clerc}
        \affiliation{Departamento de F\'isica and Millennium Institute for Research in Optics, Facultad de Ciencias F\'isicas y Matem\'aticas,
        Universidad de Chile, Casilla 487-3, Santiago, Chile}
 \author{S. Coulibaly}
        \affiliation{Universit\'e de Lille, CNRS, UMR 8523--PhLAM--Physique des Lasers Atomes et Mol\'ecules, F-59000 Lille, France}
\author{M. Tlidi}
        \affiliation{D\'epartement de Physique, Facult\'e des Sciences, Universit\'e Libre de Bruxelles (U.L.B.), CP 231, 
        Campus Plaine, B-1050 Bruxelles, Belgium}
\begin{abstract}
 We show analytically and numerically  that time delayed nonlocal response induces traveling localized states in bistable systems. 
 These states result from fronts interaction. We illustrate this mechanism in a generic bistable model with a nonlocal delayed response. 
Analytical expression of the width  and the speed of  traveling localized states are derived.
Without time delayed nonlocal response traveling localized states are excluded. 
 Finally, we consider  an experimentally relevant system, the fiber cavity with the non-instantaneous  Raman response, and show evidence  
 of traveling localized state. In addition, we propose realistic parameters and perform numerical simulations of the governing model equation.    
        \end{abstract}

 \maketitle
        
        Macroscopic systems  are  regularly described by a small number of coarse-grained or macroscopic variables. 
 The separation of time scales makes this reduction possible. Generally speaking, it allows for a description 
 in terms of slowly varying macroscopic or averaging variables \cite{Ehrenfest,Gorban,Voth}. 
 When  scales separation of the micro- and macroscopic variables 
 is not well established, the dynamics can be altered by the effect  of temporal correlations 
 that generate a temporal delayed response.
A classic example  is the polarization or the magnetization of a 
material subjected to external electric or magnetic fields. For low field strengths, 
at any given time, the polarization is ${\bf P}(\tau)=\int_{-\infty}^\tau \chi(\tau-\tau^\prime){\bf E}(\tau^\prime)d\tau^\prime$ with ${\bf E}(t)$ 
is the electric field \cite{Landau}. 
The electrical linear susceptibility $\chi (\tau)$ is the temporal delay kernel that accounts for the 
response of the charge density to the electrical stimulus. When the characteristic frequency of the electric field  is much smaller than the frequency associated with the motion 
of the electric charges, the effect of time delay can reasonably be neglected, and  the response of the system is instantaneous and local. However, nonlocal response is the rule rather than the exception and it is relevant not only for optical and magnetic systems such as  scattering of light in a continuous 
medium  \cite{Landau},  nonlinear optics \cite{Bloembergen}, and fiber optics \cite{Agrawal}
 but also in other branches on nonlinear science such as  population dynamics  \cite{Murray2001}.   
 In this case,  the birth rate includes the effect of the maturation of the population.

In another line of research, frequency comb generation in microresonators has witnessed tremendous progress 
in the last years, allowing new applications in  metrology and spectroscopy \cite{Kippenberg}.  
Frequency combs generated in optical Kerr resonators are nothing but
the spectral content of the stable temporal localized structure (LS) occurring in the cavity \cite{Scroggie,LEO}.  These peaks are generated close to the modulational instability.
The link between the phenomenon of Kerr comb formation and temporal localized structures has been established  
(see the latest overview \cite{Lugiato_2018} and reference therein). 
In optical fibers, the nonlocal delayed response is provided by the Raman effect which occurs spontaneously when an intense 
optical beam is passed through a fiber.  It has been recently shown, that Raman effect may induce 
Kerr optical frequency comb generation \cite{Cherenkov2017}. 
So far, however, the formation of traveling localized structures induced by time delayed response,  including the Raman effect, 
have neither been experimentally determined nor theoretically predicted. 
We address here the theoretical side of this problem in the regime devoid of any modulational instability.         
        
 In this letter, we show that the nonlocal delay induces traveling localized states in a generic bistable system. 
 We provide an analytical understanding of the generation of traveling LS in terms of fronts interaction.  
 We characterize these moving structures  by  deriving a coupled equations for the slow time evolution 
 of their width and their speed. Numerical simulations show a fairly good agreement 
 with the theoretical predictions. Finally, as an application, we consider all optical fiber cavity  with a
nonlocal delayed response, modeled by the Raman effect. 
We demonstrate numerically by using realistic parameters  that this simple optical device supports traveling LS. 
 
We consider the  generic bistable system with a delayed nonlocal response
        \begin{equation}
        \partial_t u=\eta+\mu u-u^3+D \partial_{\tau \tau}u+\int_{-\infty}^\tau \chi (\tau-\tau^\prime)u(\tau^\prime) d\tau^\prime,
        \label{Eq-BiestableModel}
        \end{equation}  
        where $u=u(t,\tau)$ is the scalar order parameter, $t$ and $\tau$ account, respectively, for the  slow and fast temporal evolutions. 
$\mu$ and $\eta$ are the bifurcation parameters, $D$ is the dispersion coefficient, 
and   $\chi (\tau)$ is the delay kernel function. For the sake of simplicity, we consider an exponential kernel
         \begin{equation}
         \chi (\tau)=\frac{\gamma}{\alpha} e^{\alpha \tau},
        \end{equation}  
        where $\gamma$ and $\alpha$ account for the strength of the nonlocal  delayed response and the characteristic correlation time, respectively. 
  \begin{figure}[t]
        \includegraphics[width=.47\textwidth]{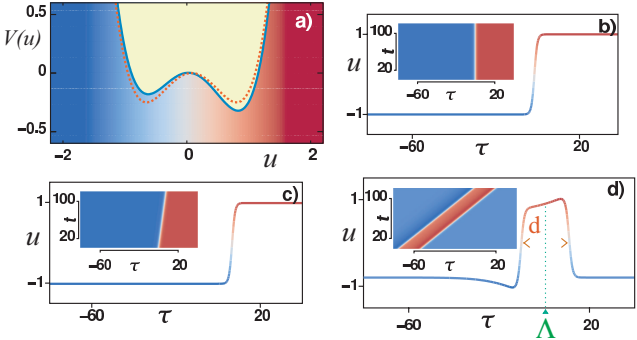} 
        \caption{ (color online)   Fronts and Traveling localized structure of Eq.~(\ref{Eq-BiestableModel}).
        (a) The potential $V\left(u\right)$ obtained for  $\gamma=0$, 
        $\mu=1$, $\eta=0$ (dashed line), and $\eta=-0.03$ (solid line). 
        (b) The front solution at the Maxwell point. Parameters are $\gamma=0$, $\mu=1$, and $\eta=0$. 
        (c) The front propagation  obtained for  $\gamma=0$, $\mu=1$, and $\eta=-0.03$. 
        (d) The traveling localized state in presence of the nonlocal delayed response  $\gamma=-0.4$, $\alpha=0.1$, $\mu=1$, and $\eta=-0.03$. 
        All profiles are obtained from the numerical integration of Eq.~(\ref{Eq-BiestableModel}). 
        The insets in (b), (c) and (d) account for $t-\tau$ maps.}
        \label{Fig1-BiestableModel}
        \end{figure}
             In the absence of nonlocal delay i.e, $\gamma = 0$, Eq.~(\ref{Eq-BiestableModel}) can be written as 
$\partial_t u=-\partial_u V(u)+ D\partial_{\tau\tau}u$ with  $V\left(u\right)=-\eta u-\mu u^2/2+u^4/4$. For $\eta^2 < \mu^{3}/9$, 
the potential $V(u)$ possesses two symmetric front solutions which are motionless only at the 
Maxwell point defined by  $\eta=0$ as shown by the dashed curve of Fig.~\ref{Fig1-BiestableModel}a. 
The motionless front connecting the two symmetric states; $u_0=\pm \sqrt{\mu}$; is plotted in 
Fig.~\ref{Fig1-BiestableModel}b. For $\gamma = 0$, and at the Maxwell point, Eq.~(\ref{Eq-BiestableModel}) 
admits an exact nonlinear front solution given by $u_\pm(\tau,\tau_p)=\pm\sqrt{\mu} \tanh[\sqrt{\mu/2}(\tau-\tau_p)]$, with  $\tau_p \equiv \int_{-\infty}^{\infty} \tau\partial_{\tau} u_\pm(\tau)d\tau /\int_{-\infty}^{\infty} \partial_{\tau}u_\pm(\tau)d\tau$ 
is the position of the front. Far from the Maxwell point $\eta \neq 0$ (solid line in Fig.~\ref{Fig1-BiestableModel}a), 
the front exhibits a motion with a constant speed as shown in Fig.~\ref{Fig1-BiestableModel}c.  When two opposite fronts are at some distance from each other, they interact in an attractive or repulsive way depending on the sign  of $\eta$. 
Front interaction was widely  used to describe motionless localized  structure formation in spatially extended  systems \cite{Coullet2002}.
In the absence of nonlocal delay traveling LS are excluded. 
However, when taking into account the delayed nonlocal response, moving LS persists for long time evolution. An example of 
such behavior is shown in Fig.~\ref{Fig1-BiestableModel}d. The homogeneous equilibria of  Eq.~(\ref{Eq-BiestableModel}); $u_0$; satisfies $\eta=-(\mu+\gamma)u_0+u_0^3$, are altered by the nonlocal delay effect. Let us consider the superposition of two well-separated fronts as 
        \begin{equation}
        u=u_+(\tau-\Lambda+d/2)  +u_-(\tau-\Lambda-d/2) 
        -\sqrt{\mu} +w.
        \label{Eq-FrontInteractionAnsatz}
        \end{equation} 
Where $d=d(t)$ and $\Lambda=\Lambda(t)$ account for the width and centroid between well separated fronts ($d(t) \sqrt{\mu}\gg 1$) as indicated in Fig.~\ref{Fig1-BiestableModel}d. We add to the superposition of the two fronts, a small perturbation  $w=w(\tau, \Lambda, d) $ with $w\ll 1$.    Replacing the ansatz (\ref{Eq-FrontInteractionAnsatz}) in Eq.~(\ref{Eq-BiestableModel}), linearizing in $w$, and 
        imposing the solvability condition after straightforward calculations, we obtain
        \begin{eqnarray}
        \dot{d}&=& f -A e^{-\sqrt{2 \mu}d}+B e^{- \alpha d},
        \label{eq-FrontInteraction} \\
        \dot{\Lambda}&=&C,
        \label{Eq-FrontInteractionDelta} 
        \end{eqnarray} 
        where 
        \begin{eqnarray}
        f&\equiv& \frac{4 \sqrt{\mu}}{||\psi||^2} \eta,  \quad { \mbox{with }}        ||\psi||^2 \equiv \int_{-\infty}^{\infty}  (\partial_\tau u_+ -\partial_\tau u_-)^2 d\tau, \nonumber \\ 
        A&\equiv&\frac{6 \mu \int_{-\infty}^{\infty} e^{-\sqrt{2 \mu} \tau} \partial_\tau u_+ d\tau }
        {||\psi||^2 }, \nonumber \\ 
        B&\equiv&\frac{2 \sqrt{\mu}  \int_{-\infty}^{\infty} e^{-\sqrt{\alpha} |\tau|} \partial_\tau u_+ d\tau }
        {||\psi||^2 } \gamma,
        \nonumber \\
         C&\equiv&-\frac{2\sqrt{\mu} \int_{-\infty}^{\infty} e^{-\alpha |\tau|} \partial_\tau u_+ d\tau }
        {\int_{-\infty}^{\infty}  (\partial_\tau u_+ +\partial_\tau u_-)^2 d\tau} \gamma .   \nonumber
        \end{eqnarray} 
 \begin{figure}[t]
        \includegraphics[width=.49\textwidth]{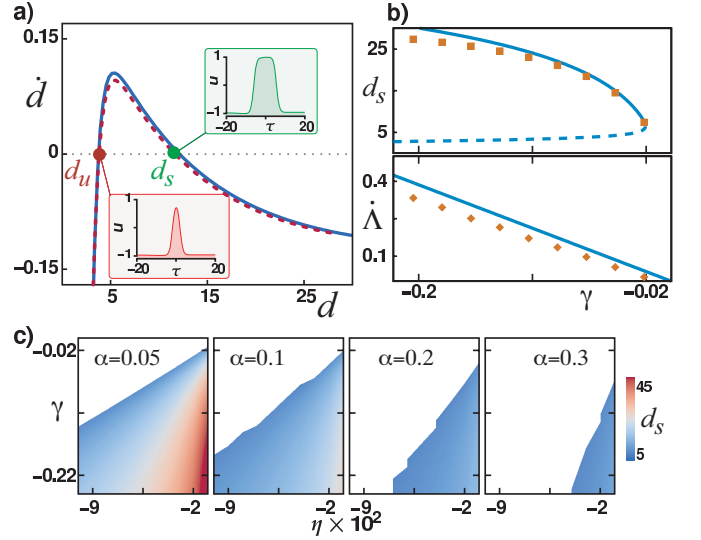} 
        \caption{ (color online) Fronts interaction. (a) Interaction force $\dot{d}$ obtained by plotting  
        Eq.~(\ref{eq-FrontInteraction}) as a function of the width $d$ (solid curve). 
        The dotted curve obtained from numerical simulations of Eq.~(\ref{Eq-BiestableModel}). Both curves are obtained for 
        $\eta=-0.03$, $\mu=1.00$, $\gamma=-0.10$, and $\alpha=0.10$. 
        The top and bottom insets correspond to the stable $d_s$ and unstable $d_u$ traveling localized state. 
        (b) The width $d_s$  (top) and speed  $\dot \Lambda$ (bottom) of the traveling  localized structure as a function of the strength of delayed nonlocal response
        $\gamma$. Continuous (dashed) curve corresponds to stable (unstable) traveling localized 
        structure obtained analytically  from Eqs.~(\ref{eq-FrontInteraction}) and (\ref{Eq-FrontInteractionDelta}).
        Dot points obtained from numerical simulations of Eq.~(\ref{Eq-BiestableModel}).
         (c) Stability domain of traveling localized structures  in the plane $\left(\eta,\gamma\right)$ 
         for different values of the correlation time $\alpha$. The colormap corresponds to traveling localized structures width. }
        \label{Fig2-frontInteraction}
\end{figure}
        Equations (\ref{eq-FrontInteraction}) and (\ref{Eq-FrontInteractionDelta}) describe  the interaction between fronts. 
        The term $f$ is proportional to $\eta$. It describes the propagation of fronts due to the potential difference 
        between the uniform states.     The term proportional to $A>0$ describes the interaction between fronts originated 
        from the superposition of their  tails. This contribution to the fronts interaction is  always attractive. 
The time-delayed response  adds a new contribution proportional to $B$ in the interaction between fronts.  
This term, however, can be either attractive for $\gamma> 0$, or   repulsive if $\gamma> 0$. Equation  (\ref{eq-FrontInteraction}) 
possesses two equilibria. 
The first  one; $d_u$; corresponds to traveling LS with small width, is always unstable. The other equilibrium with larger width; $d_s$;  is stable as shown in  Fig.~\ref{Fig2-frontInteraction}a.  The profiles of traveling LS solutions are depicted in the insets of this figure.  From dynamical system theory,  these traveling LS appear thanks to a saddle-node bifurcation as shown in Fig.~\ref{Fig2-frontInteraction}b. They are asymmetric solutions of Eq.~(\ref {Eq-BiestableModel}) (see also Figure~\ref{Fig1-BiestableModel}d).  This asymmetry is contained in the adjustment function $w$, and it is originated from the nonlocal delayed response. 
Figure~\ref{Fig2-frontInteraction}c shows the stability domain of traveling LS in the plane $\left(\eta,\gamma\right)$ 
by varying the characteristic correlation time associated with the delayed nonlocal response.  
The analytical results are compared with numerical simulations of Eq. ~(\ref {Eq-BiestableModel}). 
This comparison involves characteristics of traveling LS such as the width ($d$) and the speed ($\dot{\Lambda}$)
as a function of the strength of nonlocal delay, $\mu$, and $\eta$. In a good agreement with the analytical predictions, the width of traveling LS increases with 
the strength of all parameter values except for $\eta$.  

In brief,  the stabilization mechanism of traveling LS is attributed to nonlocal delayed response in the form of exponential.
In addition, the same mechanism is applied to others types of kernels such as Gaussian, Lorentzian, bounded support, and so forth.

        \begin{figure}[b]
        \centering{ \includegraphics[width=.45\textwidth]{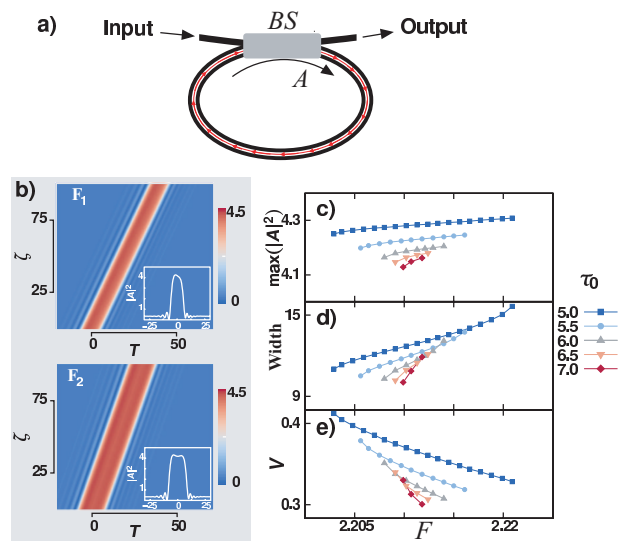} 
        \caption{ (color online)   Traveling localized structures in all fiber cavity with Raman response.  
        (a) Schematic representation of an  optical fiber ring cavity. BS accounts for the beam splitter.
        (b) $\zeta-T$ maps of traveling localized structures  of Eq.~(\ref{Eq-DDNLS}) obtained for $\Delta=4$, $\tau_0=5$, $F_1=2.202$,   ($F_2=2.223$),
         $f_R=0.18$, $\tau_1=0.012$,  and $\tau_2=0.032$.  (c) The maximum peak intensity, (d) the width, and (e) the 
         speed  of the traveling localized structures  as a function 
         of driven field amplitude as obtained for the same parameter as in (b) except for $F$ and $\tau_0$.}
        \label{Fig0-FiberCavity}}
        \end{figure}
        
In what follows, we shall apply the above mechanism to a realistic  and relevant physical system.  
For this purpose,  
we consider all fiber cavity driven coherently by an injected light beam as shown in  
Fig.~\ref{Fig0-FiberCavity}a. 
The envelope of the electric field within the cavity  is described by the Lugiato-Lefever equation (LL) \cite{LugiatoLefever} 
with the Raman delayed nonlocal response  \cite{Chembo2015}
        \begin{eqnarray}
        \frac{\partial A}{\partial \zeta}&=& F-(1+i\Delta) A-i b_2\frac{\partial^2A}{\partial T^2}+i \left(1-f_R\right)\left|A\right|^2A\nonumber \\
        &+&i f_R A \int_{-\infty}^T h^\prime_R(T-T^\prime)|A(T^\prime)|^2dT^\prime.
        \label{Eq-DDNLS}
        \end{eqnarray}
 The slow time describing the evolution over the successive round trips is $t=t_R \zeta/\kappa$. 
 The time $\tau=\tau_0T$ with $\tau_0=\sqrt{\left|\beta_2L\right|/(2\kappa)}$ is the fast time 
 in the reference frame moving with the group velocity of the light within the cavity. 
 The injected field is  $E_{in}=\sqrt{\kappa^3/(\Gamma L\theta)}F$, $\theta$ effective transmission of the beam splitter, 
 the intracavity field is $E\left(t,\tau\right)=\sqrt{\kappa/(\Gamma L)} A\left(\zeta,T\right)$, and the normalized detuning is $\Delta=\delta/\kappa$. 
 The losses, the phase detuning, the chromatic dispersion coefficient, the cavity length, the Kerr nonlinear coefficient, 
 and the  cavity roundtrip time are denoted by $\kappa$, $\delta$,  $\beta_2$, $L$, $\Gamma$, and $T_R$, respectively.   
 The coefficient $b_2$ is positive assuming normal dispersion regime, and it can be scaled down to 
 unity. Finally,  the Raman response is modeled by the function  
$h^\prime_R\left(T\right)=\tau_0 h_R\left(\tau_0 T\right)$, and  $f_R$ measures the strength of the Raman response. In agreement with experiments, a more accurate form of this function has been proposed in \cite{Agrawal_OL_2006}
         \begin{equation}
        h_R(\tau)=\frac{\tau_1^2+\tau_2^2}{\tau_1\tau_2^2} e^{\tau/\tau_2}\sin\left(\tau/\tau_1\right).
        \end{equation}  
In the absence of the Raman effect ($f_R=0$), we recover, the well known Lugiato-Lefever equation, 
that constitutes the paradigmatic  model for the study of dissipative structures in nonlinear optics  \cite{LugiatoLefever} 
(see also a recent special issue on that model,  \cite{LugiatoLafever2017} and reference therein).  In absence of the Raman effect, i.e. $f_R=0$, 
fronts  \cite{Coen}, motionless LS connecting homogeneous solutions  \cite{Parra-Rivas}, and 
traveling patterns under the effect of a walk-off \cite{walgraeraf}, delayed feedback \cite{delay_motion}, and  in the presence of the third order dispersion 
\cite{BahloulCherbi} have been reported.

As expected, the Raman effect may induce traveling  LS in an optical cavity.  
Indeed,  examples of traveling LS obtained from numerical simulations  of 
Eq.~(\ref{Eq-DDNLS}) are shown in Fig.~\ref{Fig0-FiberCavity}b.  
These structures have a fixed intrinsic width for a fixed value of parameters.   
They are asymmetric nonlinear objects with a pronounce oscillatory tails (cf. the insets of Fig.~\ref{Fig0-FiberCavity}b). 
The  characteristics of these solutions such as  width, maximum peak intensity, and speed are strongly affected by the driving field amplitude 
and  by the group velocity dispersion through the characteristic time $\tau_0$ as shown in Figs.~\ref{Fig0-FiberCavity}c, \ref{Fig0-FiberCavity}d, 
and \ref{Fig0-FiberCavity}e. From these figures, we see that the existence domain of moving LS increased 
with $ \tau_0$, and hence with the group velocity dispersion. The maximum peak intensity, 
as well as the width of moving LS, increased with the injected field amplitude as shown in Figs.~\ref{Fig0-FiberCavity}c and \ref{Fig0-FiberCavity}d. 
However, when increasing the injected field amplitude, the speed of traveling solutions decreases as shown in  Fig.~\ref{Fig0-FiberCavity}e.

        Due to the universal nature of the model Eq.~(\ref{Eq-BiestableModel}), 
 one expects that any bistable system 
with nonlocal delay  may be able to generate traveling LS.       
A physical system that meets these conditions is  the fiber optic ring cavities or microresonator comb generator. 
We now provide possible experimental parameter values relevant to the observation of traveling LS in all fiber cavity. 
We have assumed a fiber ring cavity with $L=10$ m, $\beta_2\in[10^{-6};1.8\cdot10^{-6}]$ ps$^2$/m, a finesse $\mathcal{F}=\pi/\alpha=12$,
$f_R=0.18$, $\tau_1=0.012$,  and $\tau_2=0.032$.

In conclusion, we have shown analytically and numerically that nonlocal delayed response  stabilizes 
traveling  localized structures in a generic bistable model. We characterized these solutions by computing their 
width and their speed. Without a nonlocal response, these solutions are excluded.    
Numerical solutions of the governing equations are in close agreement with 
analytical predictions. In the last part, we have considered a realistic model describing 
all fiber cavity where the nonlocal delayed coupling corresponds to the non-instantaneous  Raman response. 
This simple optical device supports traveling localized structures.  
This result strongly contrasts with those of previous studies and opens up new possibilities for the observation of traveling localized structures in practical systems.

The authors acknowledge  A. Mussot for fruitful discussions. 
This research was funded by Millennium Institute for Research
 in Optics (MIRO) and FONDECYT projects 1180903.  M. T. received support from 
the Fonds National de la Recherche Scientifique (Belgium). 
S.C. acknowledges the LABEX CEMPI (ANR-11-LABX-0007) 
as well as the Ministry of Higher Education and Research, 
Hauts de France council and European Regional Development Fund (ERDF) 
through the Contract de Projets Etat-Region (CPER Photonics for Society P4S).
      

\begin{references}
        %
        %
        \bibitem{Ehrenfest} P. Ehrenfest, T. Ehrenfest-Afanasyeva, {\it The Conceptual Foundations of the Statistical
       Approach in Mechanics. In: Mechanics} Enziklop\"adie der Mathematischen
        Wissenschaften, vol. 4. (Leipzig 1911).
        
        	
        
        \bibitem{Gorban}  {\it Model reduction and coarse-graining approaches for multiscale phenomena},
        edited by A.N. Gorban, N.K. Kazantzis, I.G. Kevrekidis, H.C. Ottinger, and C. Theodoropoulos, 
        (Springer-Verlag Berlin Heidelberg, 2006).
        
        \bibitem{Voth}  G. A. Voth,   {\it. Coarse-graining of condensed phase and biomolecular systems}, (CRC press, Boca Raton, 2008).
        
        
        \bibitem{Landau}  L.D. Landau, and E.M. Lifshitz, {\it Course of Theoretical Physics. Vol. 8: Electrodynamics of Continuous Media},
        (Oxford, 1960).
        
        \bibitem{Bloembergen}  N. Bloembergen,  {\it Nonlinear optics} (World Scientific, 1996).
        
                \bibitem{Agrawal} G.P. Agrawal, {\it Nonlinear fiber optics}  (Springer, Berlin, Heidelberg, 2000).
        
        \bibitem{Murray2001} J.D. Murray, {\it Mathematical Biology} 
       (Springer-Verlag, New York, 2001).
       
       \bibitem{Kippenberg} P. Del'Haye,   A. Schliesser, O. Arcizet, T. Wilken, R. Holzwarth, and T. J. Kippenberg,  
       Nature, {\bf 450}, 1214 (2007). 
       
       

        
        \bibitem{Scroggie} A. J. Scroggie, W. J. Firth,  G. S. McDonald, M. Tlidi, R., Lefever,  and L. A. Lugiato,   
         Chaos, Solitons \& Fractals, {\bf 4}, 1323 (1994).
         
         \bibitem{LEO}  F. Leo,  S. Coen, P.  Kockaert, S.P. Gorza, P. Emplit,  and M. Haelterman,  Nature Photonics, {\bf 4}, 471 (2010).
         
             \bibitem{Lugiato_2018}L.A. Lugiato, F. Prati, M.L. Gorodetsky, and T.J. Kippenberg,  Phil. Trans. R. Soc. A, {\bf 376}, 20180113 (2018).   
       
         \bibitem{Cherenkov2017} A.V. Cherenkov,  N.M. Kondratiev, V.E. Lobanov, A.E. Shitikov,  D.V. Skryabin,  
        and  M.L. Gorodetsky,  Opt. express, {\bf 25}, 31148 (2017).
        
         \bibitem{Coullet2002} P. Coullet,   Int. J. Bifurcation Chaos, {\bf 12}, 2445 (2002); M. G. Clerc, and C. Falcon, 
        Physica A {\bf 356}, 48 (2005); C. Fernandez-Oto, M.G. Clerc D. Escaff, and M. Tlidi, Phys. Rev. Lett. {\bf 110}, 174101 (2013);
        D . Escaff, C. Fernandez-Oto, M.G. Clerc, and M. Tlidi, Phys. Rev. E {\bf 91}, 022924 (2015).
        
        
         \bibitem{LugiatoLefever} L. A. Lugiato and R. Lefever,  Phys. Rev. Lett. {\bf 58}, 2209 (1987).
         
        \bibitem{Chembo2015}  Y. K. Chembo, I. S. Grudinin, and N. Yu, Phys. Rev. A {\bf 92}, 043818 (2015).
        
         \bibitem{Agrawal_OL_2006}  Q. Lin and G.P.  Agrawal,  Optics letters {\bf 31}, 3086 (2006).    
         
                 \bibitem{LugiatoLafever2017}  Y.K. Chembo, D. Gomila, M. Tlidi, and C.R. Menyuk,  
        Theory and applications of the Lugiato-Lefever Equation,
        Eur. Phys. J. D {\bf 71}, 299 (2017).
         
        
        \bibitem{Coen} S. Coen, M. Tlidi, P. Emplit, and M. Haelterman, Phys. Rev. Lett. {\bf 83}, 2328 (1999).
        
  
        
        \bibitem{Parra-Rivas} P. Parra-Rivas, E. Knobloch, D. Gomila, and L. Gelens
Phys. Rev. A {\bf 93}, 063839 (2016).


\bibitem{walgraeraf}M. Santagiustina, P. Colet, M. San Miguel, D. Walgraef,  Physical review letters 79, 3633 (1997).
  \bibitem{delay_motion} K. Panajotov, D. Puzyrev, A.G. Vladimirov, S.V. Gurevich, M. Tlidi, Phys. Rev. A, 93, 043835 (2016).
  
     \bibitem{BahloulCherbi} M.Tlidi, L. Bahloul, L. Cherbi, L.,  A. Hariz, and S. Coulibaly, 
     Phys. Rev. A, {\bf 88}, 035802 (2013).
        
        
 
 \end{references}

 \end{document}